\documentclass{emulateapj}

  \usepackage{natbib}
  \bibpunct{(}{)}{;}{a}{}{,} 

\newcommand{\degree}{\ensuremath{^\circ}}

\shorttitle{The NIR sky brightness at Calar Alto}
\shortauthors{S\'anchez et al.}

\begin{document}

\title{The night-sky at the Calar Alto Observatory II: The sky at the near infrared}

\author{S.F~S\'anchez\altaffilmark{1}, 
U.~Thiele\altaffilmark{1},
J.~Aceituno\altaffilmark{1},
D.~Cristobal\altaffilmark{2},
J.~Perea\altaffilmark{3},
J.~Alves\altaffilmark{1},
}

\altaffiltext{1}{Centro Astron\'omico Hispano Alem\'an, Calar Alto, (CSIC-MPG),
  C/Jes\'us Durb\'an Rem\'on 2-2, E-04004 Almeria, Spain }
\altaffiltext{2}{Instituto de Astrof\'\i sica de Canarias, Via Lactea S/N, La
  Laguna, S/C de Tenerife (Spain)}
\altaffiltext{3}{Instituto de Astrof\'\i sica de Andalucia, Camino Bajo de
  Huetor S/N, Granada (Spain)}

\email{sanchez@caha.es}

\begin{abstract}
   
  We present the characterization of additional properties of the night-sky at
  the Calar Alto observatory, following the study started in S\'anchez et
  al. (2007), hereafter Paper I. We focus here on the night sky-brightness at
  the near-infrared, the telescope seeing, and the fraction of useful time at
  the observatory. For this study we have collected a large dataset comprising
  7311 near-infrared images taken regularly along the last four years for the
  ALHAMBRA survey ($J$, $H$ and $Ks$-bands), together with a more reduced
  dataset of additional near-infrared images taken for the current study. In
  addition we collected the information derived by the meteorological station
  at the observatory during the last 10 years, together with the results from
  the cloud sensor for the last $\sim$2 years. We analyze the dependency of
  the near-infrared night sky-brightness with the airmass and the seasons,
  studying its origins and proposing a zenithal correction. A strong
  correlation is found between the night sky-brightness in the $Ks$-band and
  the air temperature, with a gradient of $\sim -$0.08 mag per 1 $\degree$ C.
  The typical (darkest) night sky-brightness in the $J$, $H$ and $Ks$-band are
  15.95 mag (16.95 mag), 13.99 mag (14.98 mag) and 12.39 mag (13.55 mag),
  respectively. These values have been derived for the first time for this
  observatory, showing that Calar Alto is as dark in the near-infrared as
  most of the other astronomical astronomical sites in the world that we could compare
  with. Only Mauna Kea is clearly darker in the
  $Ks$-band, but not only compared to Calar Alto but to any other observatory
  in the world. The typical telescope seeing and its distribution was derived
  on the basis of the FWHM of the stars detected in the considered
  near-infrared images. This value, $\sim$1.0$\arcsec$ when converted to the
  V-band, is only slightly larger than the atmospheric seeing measured at the
  same time by the seeing monitor, $\sim$0.9$\arcsec$. Therefore, the effects
  different than the atmosphere produce a reduced degradation on the telescope
  seeing, of the order of $\sim$10\%. Finally we estimate the fraction of
  useful time based on the relative humidity, gust wind speed and presence of
  clouds. This fraction, $\sim$72\%, is very similar to the one derived in
  Paper I, based on the fraction of time when the extinction monitor is
  working.

\end{abstract}



\section{Introduction}

The night sky brightness in the optical and near-infrared, the number
of clear nights, the seeing, transparency and photometric stability
are some of the most important parameters that qualify a site for
front-line ground-based astronomy. There is limited control over all
these parameters, and only in the case of the sky brightness is it
possible to keep it at its natural level by preventing light pollution
from the immediate vicinity of the observatory. Previous to the
installation of any observatory, extensive tests of these parameters
are carried out in order to find the best locations, maximizing then
the efficiency of these expensive infrastructures.  However, most of
these parameters are not constant along the time.

We have started a program to determine the actual values of the main
observing conditions for the Calar Alto observatory.  The Calar Alto
observatory is located at 2168m height above the sea-level, in the
Sierra de los Filabres (Almeria-Spain) at $\sim$45 km from the
Mediterranean sea. It is the second largest European astronomical site
in the north hemisphere, behind the Observatorio del Roque de los
Muchachos (La Palma), and the most important in the continental
Europe. Currently there are six telescopes located in the complex,
three of them operated by the Centro Astronomico Hispano Aleman
A.I.E. (CSIC-MPG), including the 3.5m, the largest telescope in the
continental Europe.

Along its 26 years of operations there has been different attempts to
characterize some of the main properties described before: (i) Birkle et
al. (1976) presented the first measurements of the seeing at Calar Alto, being
the basis of the site testing; (ii) Leinert et al. (1995) determined the
optical sky brightness corresponding to the year 1990; (iii) Hopp \& Fernandez
(2002) studied the extinction curve corresponding to the years 1986-2000; (iv)
Ziad et al. (2005) estimated the median site seeing in the observatory from a
single campaign in may 2002. A consistent study of all these properties was
presented in S\'anchez et al. (2007), including the first results of our
program.

In that study, hereafter Paper I, we focused on the optical properties of the
night sky, presenting (i) the first optical night-sky spectrum, identifying
the natural and light pollution emission lines and their strength, (ii) the
moon-less night-sky brightness in different optical bands, (iv) the extinction
and its yearly evolution and (v) the atmospheric seeing and its yearly
evolution. It was found that most of these properties were similar to those of
major astronomical sites. In particular Calar Alto is among the darkest places
in the world in the optical range.

In the current study we focus on the near infrared (NIR) properties of the sky
at Calar Alto. In particular we characterize here the typical and darkest sky
brightness in the $J$, $H$ and $K$-bands, comparing them with similar
properties in other observatories world-wide. Their dependencies with
positioning in the sky and seasons are also analyzed. The typical seeing
measured in real observations (telescope seeing) is derived and compared with
the atmospheric seeing estimated by the seeing monitor (out of the
telescope). Finally we estimate the fraction of useful time in the observatory
based on the data obtained by the meteorological station along the last ten
years (relative humidity and gust wind) and the presence of clouds based on
the cloud sensor.

The structure of this article is as follows: in Section \ref{data} we describe
the dataset collected for the current study, including a description of data
and the data reduction; in Section \ref{ana} we show the analysis and results
performed over the different types of data and the results derived for each
one; in Section \ref{conc} we summarize the main results and present the
conclusions. All the magnitudes listed in this article are in the Vega
system. 

\section{Description of the Data}
\label{data}

The main purpose of this study is the characterization of the sky brightness
in the near infrared at the Calar Alto observatory. In order to address this
goal we collected different observational data, taken with different
telescopes and instruments and using different methods to calibrate, reduce
and analyze the data. With this selection we try to minimize the effects on the
results of a particular instrumental setup, maximizing the reliability of the
dataset and its time coverage.

\subsection{Observations of Calibration Stars}
\label{magic}

As we indicated in detail in Paper I the most obvious method to derive
the sky brightness is to obtain direct imaging on calibration stars
(or well calibrated fields), using the well known photometry of the
observed stars to self-calibrate the images and then derive the sky
surface brightness. With this purpose, observations were taken using
the MAGIC NIR camera (Herbst et al. 1993) at the 1.23m telescope
during the nights of the 30th and 31st of January 2008. Although a
strong dependency of the NIR sky brightness with the moon (at least
for the $H$ and $K$ bands) is not expected, both nights were selected
to be dark. In both nights we used the J, H, Ks, K and Km bands. The
MAGIC detector is an Rockwell NICMOS3 array of 256$\times$256 pixels,
with a scale of 1.2$\arcsec$/pixel and a field-of-view of
5.1$\arcmin$$\times$5.1$\arcmin$, enough for the purpose of this
study. MAGIC is under operation at Calar Alto since 1995, and it is not in
regular use anymore.

On the 30th of January we observed the core of M67, taking 24 individual
frames of 60s per band (splitted in individual integrations of 2s each
one). This cluster was selected since there are publically available NIR
photometry for most of its members and on the indicated date it remains near
the zenith (airmass$<$1.2) during most of the night. On the 31st we took 24
individual frames of 60s per band on a single calibration star from the UKIRT
faint photometric standards (Casali \& Hawarden 1992), observing it at
different airmasses. In both nights the observations were started two hours
after the twilight, in order to minize as maximum its possible effects (if
any).

The data were reduced following standard procedures for the near infrared,
using IRAF \citep{iraf}\footnote{IRAF is distributed by the National Optical
Astronomy Observatories, which are operated by the Association of Universities
for Research in Astronomy, Inc., under cooperative agreement with the National
Science Foundation.} and self programed routines. Each frame was corrected by
its corresponding dark frame and flatfielded using flux normalized
domeflats. Bad pixels were identified using both the dark and the domeflats
and masked from the individual frames. Since we are interested in the sky
brightness no sky subtraction was applied.

Once the images were reduced we measured the number of counts within an
aperture of 8$\arcsec$ for the brightest stars detected in the field (in the
case of the 30th of January data) or the photometric standard (in the case of
the 31st of January data). The number of counts from the sky was determined by
measuring it directly in an annular ring of 14$\arcsec$ and 27$\arcsec$ (inner
and outer radii). After subtracting the sky to the counts of the detected
stars the photometric zeropoint for the image was determined using the
formula:

$$\rm mag_{\rm zero} = mag_{\rm app} + ext + 2.5 {\rm log}_{\rm 10} ( counts/t_{exp})  (1)$$

where mag$_{\rm zero}$ is the magnitude zeropoint, mag$_{\rm app}$ is the
apparent magnitude of the calibration star in the corresponding band, ``ext''
is the extinction, ``counts'' are the measured counts within the indicated
aperture, and t$_{\rm exp}$ is the exposure time.  When more than one star was
analyzed we adopted the mean value of the derived zeropoints as the image
photometric zeropoint. After that the surface sky-brightness was derived for
each image by using the formula:

$$\rm SB_{sky} = mag_{zero} - 2.5 {\rm log}_{10} ( counts_{sky}/t_{exp}/scale^2) (2)$$

where SB$_{\rm sky}$ is the surface brightness in magnitudes per square arcsec,
mag$_{\rm zero}$ is the zeropoint described before, counts$_{\rm sky}$ is the sky
counts level estimated and ``scale'' is the pixel scale in arcseconds. It is
noticed that the magnitude zeropoint was corrected by the extinction, but the
sky brightness was not, following the convention adopted in most of the recent
studies of sky brightness (Paper I and references therein). Correcting the sky
brightness for extinction would be appropiate only if the extinguishing layer
were known to be below all sources of sky brightness, which is not the case
(Benn \& Ellison 1998a,b).

For the extinction we used the formulae presented in Paper I, that allows to
estimate the extinction in the NIR bands based on the $V$-band extinction from
the Calar Alto Extinction Monitor (CAVEX, PI: U.Thiele, see PaperI). In both
nights the extinction in the NIR was almost neglectible. We finally got 24
individual estimations of the surface sky brightness for each band and night.

\subsection{Observations of the ALHAMBRA survey}
\label{alh}

The ALHAMBRA survey (Moles et al. 2008) is the largest cosmological survey
currently on going at the Calar Alto Observatory. It comprises the observation
of 4 discontinuous square degrees on the sky in 8 different locations using 20
middle-width filters in the optical range and three additional ones (J, H and
Ks) in the near infrared. The locations were selected to cover areas of low
galactic extinction, without very bright stars in the field, and/or
corresponding areas already covered by other extragalactic surveys (like
COSMOS, HDFN or the GROTH-strip). The main goal of this survey is to derive
high quality photometric redshift of about $\sim$1.6 millon of galaxies, in
order to study the evolution of galaxies up to $z\sim$1. The NIR observations
of the survey are taken with the Omega2000 camera mounted in the 3.5m
telescope at Calar Alto. This camera has installed a Hawaii-2 detector with an
array of 2048$\times$2048 pixels, each one with projected size of
0.447$\arcsec$. Thus, the detector covers a field-of-view of
$\sim$15$\arcmin$$\times$15$\arcmin$. Omega2000 is under operations at Calar
Alto since 2001, and it has been for a long period of time one of the NIR
cameras with the largest field-of-view in the world. To observe the proposed
field-of-view and to achieve the proposed depth ($Ks<$20 mag, 5$\sigma$
detection limit) the ALHAMBRA survey obtains several individual frames of 80s
(J), 60s (H) and 46s (Ks) each one, at many different locations on the sky,
being an unique dataset to study the properties of the NIR sky brightness (as
an unproposed legacy).

The ALHAMBRA survey has accumulated observations with Omega2000 along 75
nights, from August 2004 to March 2008, with about $\sim$150 individual frames
per night. That means about $\sim$3500 individual frames on target per
band. These nights corresponds to a $\sim$67\% of the allocated nights for the
project in that time interval, which reinforces our claim in Paper I that a
$\sim$70\% of the nights are useful for astronomical observations at Calar
Alto (see Section 3.3 for a full discussion). 

To automate and homogenize the data reduction a semiautomatic pipeline was
created (Cristobal et al. 2008, in prep.). This pipeline performs the typical
reduction steps for NIR imaging data, including the dark-current subtraction,
flat-fielding, image combination, sky determination and subtraction and flux
calibration. The flux calibration was performed self-calibrating the images
using the NIR magnitudes of the field stars extracted from the 2MASS survey
\cite{skru06}. The typical error of the magnitudes of each individual star is
about $\sim$0.03 mag, with a range of errors between $\sim$0.02 and $\sim$0.10
mag. The large field-of-view of Omega2000 grants the detection of a
sufficiently large number of them ($>$20) to obtain an accurate flux
calibration, better than $\sim$0.025 mags (Cristobal et al., in prep.).  As a
by product of the pipeline the sky background is determined for each combined
image.

Due to its nature the different pointings of the survey are observed in
different nights, at different azimuths and elevations (and
airmasses). Therefore, the derived sky backgrounds from the combined frames
cannot be used to characterize particular nights and/or atmospheric
conditions. However, they are still useful indicators of the average sky
background in the observatory. 

For the present study we collected the night sky backgrounds available from
the ALHAMBRA reduced dataset, all of them corresponding to 12 different nights
of the same month, August 2004 (the 1st run of the survey). These data have
the advantage of their high signal-to-noise, due to the depth of the combined
images, and the fact that they have been reduced using a refined pipeline. The
final sample comprising a set of 17 individual estimations of the sky
background per filter.

A simpler procedure was developed to obtain the sky brightness for each
individual frame obtained by the survey, which comprises a much larger
dataset. For each individual frame taken on the ALHAMBRA fields the
approximate location of the field stars with NIR magnitudes derived by the
2MASS survey were obtained. Poststamp images with a field-of-view of
40$\arcsec$$\times$40$\arcsec$ around these locations were taken. In each
individual poststamp the location of the field star was determined by a simple
centroid determination, masking the pixels up to 5$\sigma$ the standard
deviation over the median background. Then, the counts within an aperture of
6$\arcsec$ radius around the determined centroid were measured, subtracting
the background counts estimated within a ring between 14$\arcsec$ and
20$\arcsec$ radius and the nominal dark-current of the instrument (Kov\'acs et
al. 2004).
Simultaneously the median counts within four different boxes of 18$\times$18
pixels located at 10$\arcsec$ away from each centroid was derived as an
estimation of the counts of the sky brightness. The estimated counts per star
were used to derive a magnitude zero-point for each star in the field with
known photometry using the formulae (1), and the corresponding sky surface
brightness using formulae (2) and the estimation of the sky counts. The
number of stars detected per field with accurate 2MASS photometry grants the
quality of the derived surface sky-brightness, despite the fact that the
accuracy of the individual photometry is not that good in a few cases, as
indicated before. The fact that no flat-field correction was applied should
not affect strongly the results due to the small variations in the efficiency
pixel-to-pixel in the Omega2000 array.

We finally got a sample of 1954, 2394 and 2963 individual estimations of the
night sky surface brightness for the $J$, $H$ and $Ks$ bands. Due to the
number of measurements and the time coverage it comprises one of the largest
samples of this nature ever obtained to our knowledge.

As a by-product of the analysis it was derived the average FWHM on each
analyzed image. For doing so it was extracted an horizontal and vertical
intensity cut along and across the estimated centroid in each poststamp
image. The intensity cuts were then fitted with a single gaussian function
plus a continuum intensity to remove the background. The average of the
horizontal and vertial FWHMs is taken as the FWHM of the image of each star,
and then, the median value of the FWHMs derived on the same image is stored. 
Once removed the effect of the pixel scale, the final FWHM can be considered
as the seeing in the detector plane in the considered band.

A summary of all the data set collected for this study is listed in Table
\ref{tab_data}, including the date of the observations or the projects from
where they were extracted, the telescope and instrument used, the bands
covered by the data and the number of individual frames analized (the number
of individual sky background estimations).


\subsection{Meteorological conditions}
\label{meteo}

The atmospheric conditions in the observatory has been continously monitorized
by a computerized meteorological station during the last decade. This station
measures the most important meteorological parameters, such as the surface
temperature, the relative humidity, the wind speed and direction, the wind
gust speed and the air preasure, and store it every 30 seconds. These data
comprises a huge dataset with 3541161 individual measurements of each
parameter for the time period between January 1998 and December 2008, that
allows to characterize the atmospheric conticions at Calar Alto.

\section{Analysis and Results}
\label{ana}


\subsection{The typical Night Sky Brightness at the near infrared}
\label{ana_sb}

The near infrared sky brightness is dominated by many instrically narrow
hydroxyl (OH) emission lines (eg., Oliva \& Origlia 1992; Rousselot et
al. 2000; Cox 2000). A few other species (eg., molecular oxygen at 1.27
$\mu$m) also contribute, as do H$_2$O lines at the long wavelength end of the
$K$-bands. Another contributions to the sky background is zodiacal emission
(eg., Harwit et al. 1969; Content 1996; Cox 2000), thermal emission from the
atmosphere and the telescope (which follows a graybody curve, Content 1996),
and the moonlight (eg., Cox 2000).

All these effects contributes differently at different wavelength ranges. In
the $J$-band the moon-light may have a strong effect, if it is observed near
the moon. On the other hand, its effect is neglectible in the $H$ and $K$. In
general, the dominant effect in the $J$ and $H$ bands is the OH emission
lines, although the effect of the zodical light is not neglectible, being
similar in both bands (eg., Content 1996). However, in the $K$-band the
thermal emission becomes the dominant effect, in particular the telescope one
(eg., Cox 2000), being less important the OH emission lines, and the optical
airglow is almost neglectible.

The night sky brightness suffers strong variations along each night which
imposse limitations to the concept of {\it typical} sky brightness. The OH
emission varies diurnally, being stronger in the daytime and at
twilight. It may decrease by a factor 2 or 3 in one or two hours after the
twilight, and it does not increase again just before the sunrise. Its emission
roughly scales with the airmass, which produces an increase of the
sky brightness at higher airmasses, expecially in the $J$ and $H$ bands.

\begin{figure}
\resizebox{\hsize}{!}
{\includegraphics[width=\hsize,angle=270]{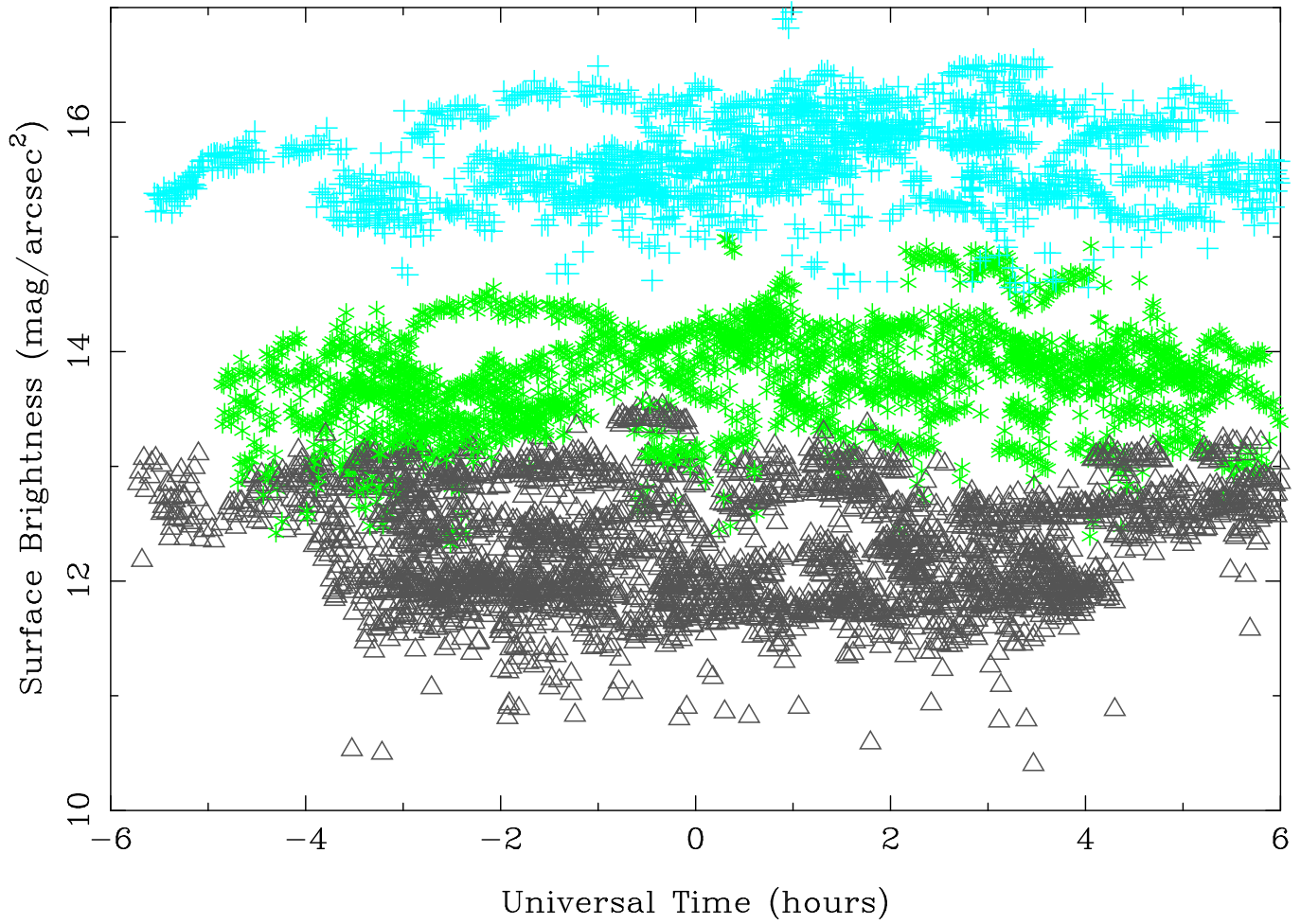}}
  \caption{\label{mag_UT} 
 Distribution of the NIR surface sky brightness in different bands, derived
 using the data from the ALHAMBRA survey, along the Universal Time, for the 75
 nights covered by our dataset, expanding from August 2004 to March 2008. The
 plot shows 1954, 2394 and 2963 individual points for the $J$, $H$ and $Ks$
 bands, respectively.  Blue crosses represent the values for the $J$-band,
 green stars represent the values for the $H$-band, and gray triangles
 represent the values for the $Ks$-band. The range of night sky surface
 brightnesses for each band are clearly appreciated.
 }
  \end{figure}

The variation in the night sky brightness due to all these effects can be
appreciated in Figure \ref{mag_UT}. This figure shows all the individual night
sky surface brightnesses that comprise the dataset derived from the ALHAMBRA
survey (described before), along the Universal Time corresponding to each of
the 75 observed nights. The wider range of surface brightness it is found in
the $Ks$-band, while the narrower one corresponds to the $J$-band.

As a rough estimation of the typical sky brightness it was derived the median
sky brightness of each data-set. For a better characterization it was also
derived the darkest sky brightness and the brightest one, which will
illustrate the strength of the variation of the sky brightness. For the full
sample of the ALHAMBRA dataset, which includes data taken at many different
epochs this latter parameter were not derived, since it turns to be
meaningless. All these estimations of the sky brightness are listed in Table
\ref{tab_med}, for each data-set and each band.

The values listed in the table illustrate how strong is the variation
of the sky brightness, not only within a certain night, but also
night-to-night. In this context the median sky brightness is the less
useful parameter to characterize the sky brightness, since it was
included in its derivation data taken at different times from the
twilight and at different airmasses. A good example of the biases that
can be found by adopting the median sky brightness are the data of the
two nights observed at the 1.23m telescope. They correspond to
observations on two consecutive nights, using the same instrument at
the same telescope. However, there is a strong difference in the
median sky brightness derived for each of these nights, at least
for the $J$ and $H$ bands. As we will show later the main reason for
these differences is due to the air-mass range of the observations
taken each night. While the 1st night the observations were restricted
almost to the zenith ($X<1.2$), in the second one it was explored a
wide range of airmasses. This may explain also the strong differences
found among the brightest sky brightness for the same bands for the
different datasets.

A more reliable parameter to characterize the sky brightness is the
darkest sky brightness derived for each band in the different
datasets. Although this parameter also presents a variation, this is
much reduced than the one found in the median sky brightness. However
it does not illustrate the expected sky brightness at the observatory
in any conditions, it just represents the darkest ever possible sky
that can be found on the data set. As expected the darkest
sky-brightness is found in the largest dataset, since it covers the
wider range of atmospheric conditions, and therefore it has the
largest probability of catching the optimal ones.

\subsubsection{Dependency on the zenith distance}

As indicated before the sky brightness presents a dependency with the zenith
distance. In the optical range this dependency is mostly due to the airglow,
an effect that is stronger at largest airmasses and smaller at the zenith (eg.,
Paper I and references therein). In the NIR there are additional components
that depend mostly on the OH emission (and the thermal emission), and which
effect is minimized also at the zenith.

In general the dependecy of the sky brightness with the airmass (ie., the
zenith distance) can be modeled by the following formula:

$$\rm \Delta m = -2.5 {\rm log}_{10}[(1-f)+f X]+\kappa (X-1)$$

The first two parameters within the logarithm of this formula are described in
Patat (2003), being $\Delta m$ the increase in sky brightness at a certain
band and airmass ($X$), $f$ the fraction of the total sky brightness generated
by airglow, and (1-$f$) the fraction produced outside the atmosphere (hence
including zodiacal light, faint stars and galaxies). Finally, $\kappa$ is the
extinction coefficient at the corresponding wavelength. As already indicated
in Paper I these two parameters describe correctly the dependency of the sky
brightness with the airmass for the optical range. Indeed the fraction $f$ is
almost constant for the optical range, being of the order of $\sim$0.6 (eg.,
Benn \& Ellinson 1998a,b). However in the near-infrared the fraction of night
sky light generated in the atmosphere can be larger than the one generated
outside the atmosphere, and therefore ``f'' can be larger than 1.

This formula is an approximation for low airmasses of the so-called van Rhijn
formula (e.g., Chamberlain 1961), which describe the increase of the airglow
emission when the zenith angle increases. The extinction correction has to be
taken into account since the light, emitted in a layer on top of the
atmosphere, is attenuated by the atmospheric extinction.

\begin{figure}
\resizebox{\hsize}{!}
{\includegraphics[width=\hsize,angle=270]{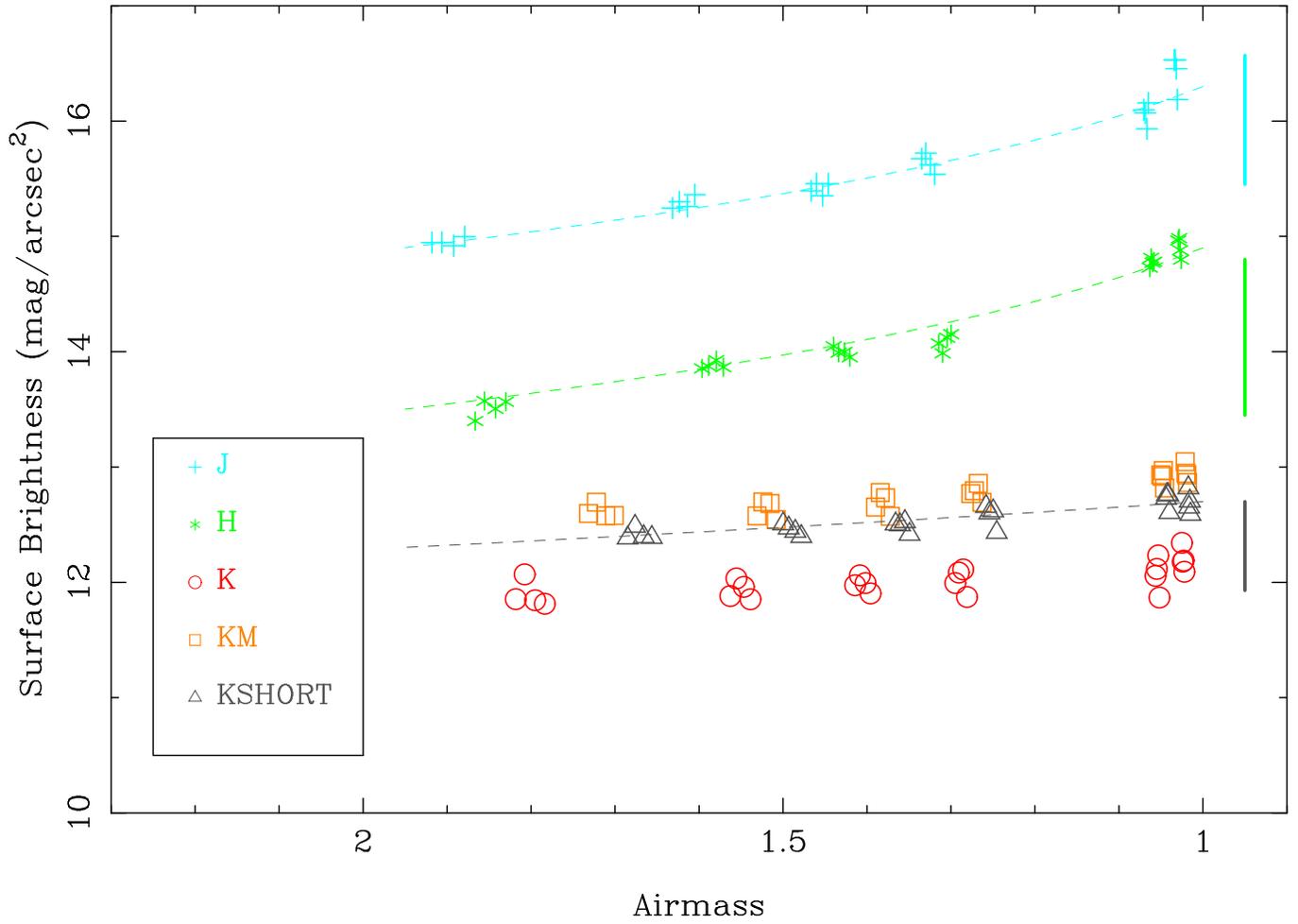}}
  \caption{\label{mag_air} 
 Distribution of the NIR surface sky brightness in different bands along the
 airmass for the night of the 31st of January 2008. The symbols for the
 different bands are described in the plot. There is clear dependency of the
 surface brightness with the airmass, particularly strong in the J and H
 bands. This dependency can be reproduced assuming an increase of the
 background flux with the airmass as it can be seen with the corresponding
 overplotted lines. A similar distribution is appreciated in all nights from
 our dataset.
 }
  \end{figure}

Figure \ref{mag_air} shows the distribution of the NIR surface sky brightness
along the airmass for the different bands observed in the night of the 31st of
January 2008. As expected, there is a clear dependency of the surface
brightness with the airmass, particularly strong in the J and H bands. A
similar result is found in any night within our different collected
dataset. This particular night was shown just as an illustrative example,
since its dataset covers the larger number of filters and a wide range of
airmasses.

The sky-brightness/air-mass distribution derived for each night and filter
were fitted with the parametrization described before. For the $J$ and $H$
band it was obtained that a value of $f\sim$2.6 reproduces well the air-mass
dependency of the night sky brightness. The similarities between both bands
are mostly due to the fact that the sky brightness is dominated in both of
them by very intense OH bands. The $J$-band covers the 7-4 and 8-5 OH bands,
while the $H$-band covers the 4-2, 5-3 and 6-4 ones (eg., Chamberlain 1961;
Maihara et al. 1993; Leinert et al. 1998). On the other hand, the air-mass
dependecy is weaker in the $K$-band. A ``f'' value of $\sim$0.4 reproduces
well its shape. The $K$-band is less affected by the OH bands. Only a part of
the 9-7 band enters at wavelengths lower than 2.2$\mu$m, and there is wide gap
without any OH band between 2.2$\mu$m and 2.7$\mu$m (eg., Leinert et
al. 1998). For this band the thermal contribution starts to be important (eg.,
Content 1996; Leinert et al. 1998; Glass 1999).

\subsubsection{The sky brightness at the Zenith}

The results from the parametrization shown in the previous section were used
to correct the sky-brightness from its dependency with the airmass and to
transform all the measured values to the corresponding value at the zenith.
As already indicated in Paper I, and references therein, these values are a
good characterization of the typical sky-brightness for an astronomical
site.

Table \ref{tab_zenit} lists the median sky-brightness for the different
considered bands after applying the considered correction for the dataset
corresponding to the full sample of ALHAMBRA data. In addition it is listed
the median values of the uncorrected sky-brightnesses restricted to the
airmasses near to the zenith ($X<1.05$) for the considered dataset. Although
this later is a much limited sample ($\sim$150 measurements), the good
agreement between both estimations of the zenithal sky-brightness, within
$\sim$0.1 mag, reinforces the validity of the adopted correction.  

\begin{figure}
\resizebox{\hsize}{!}
{\includegraphics[width=\hsize,angle=270]{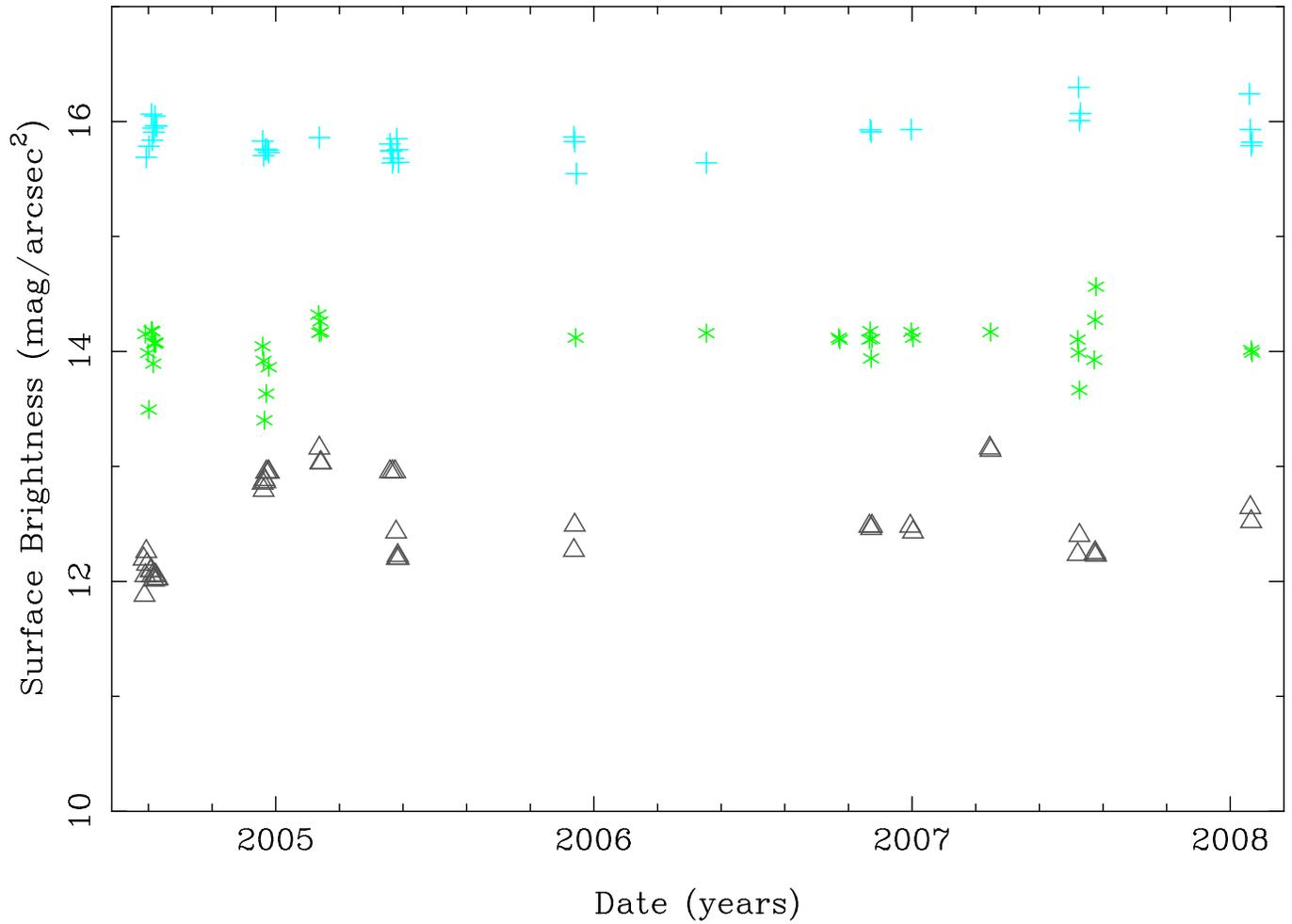}}
  \caption{\label{mag_date} Distribution of the airmass corrected NIR surface
sky brightness for the J, H and Ks bands, averaged for each night, along the
time. The symbols for each band are similar to those of the figure
\ref{mag_air}.  There is a clear seasonal trend in the Ks band, being the sky
brighter in summer than in winter.  %
}
  \end{figure}

In order to test the effects of the thermal emission on the sky-brightness the
data were splitted in two subsamples, one covering the colder epoch of the
year (from November to February, so called {\it winter}), and another
comprising the rest of the year (so called {\it summer}). The median zenithal
sky-brightness were derived for both datasets and listed in Table
\ref{tab_zenit}. Only in the $Ks$-band is appreciated a clear seasonal
difference, with the winter sky-brightness being $\sim$0.8 mags darker than
the summer one, on average. These differences in the sky brightness along the
year are clearly appreciated in Figure \ref{mag_date}, where the median
zenithal sky-brightness for each night and filter are represented along the
year. Again, only for the $Ks$-band the sky-brightness shows a clear pattern,
with a change associated with the seasons, being brighter in the summer and
darker in the winter. This result is consistent with the expected origin of
the sky-brightness at the different bands, since the thermal contribution
should affect strongly the $Ks$-band.

\begin{figure}
\resizebox{\hsize}{!}
{\includegraphics[width=\hsize,angle=270]{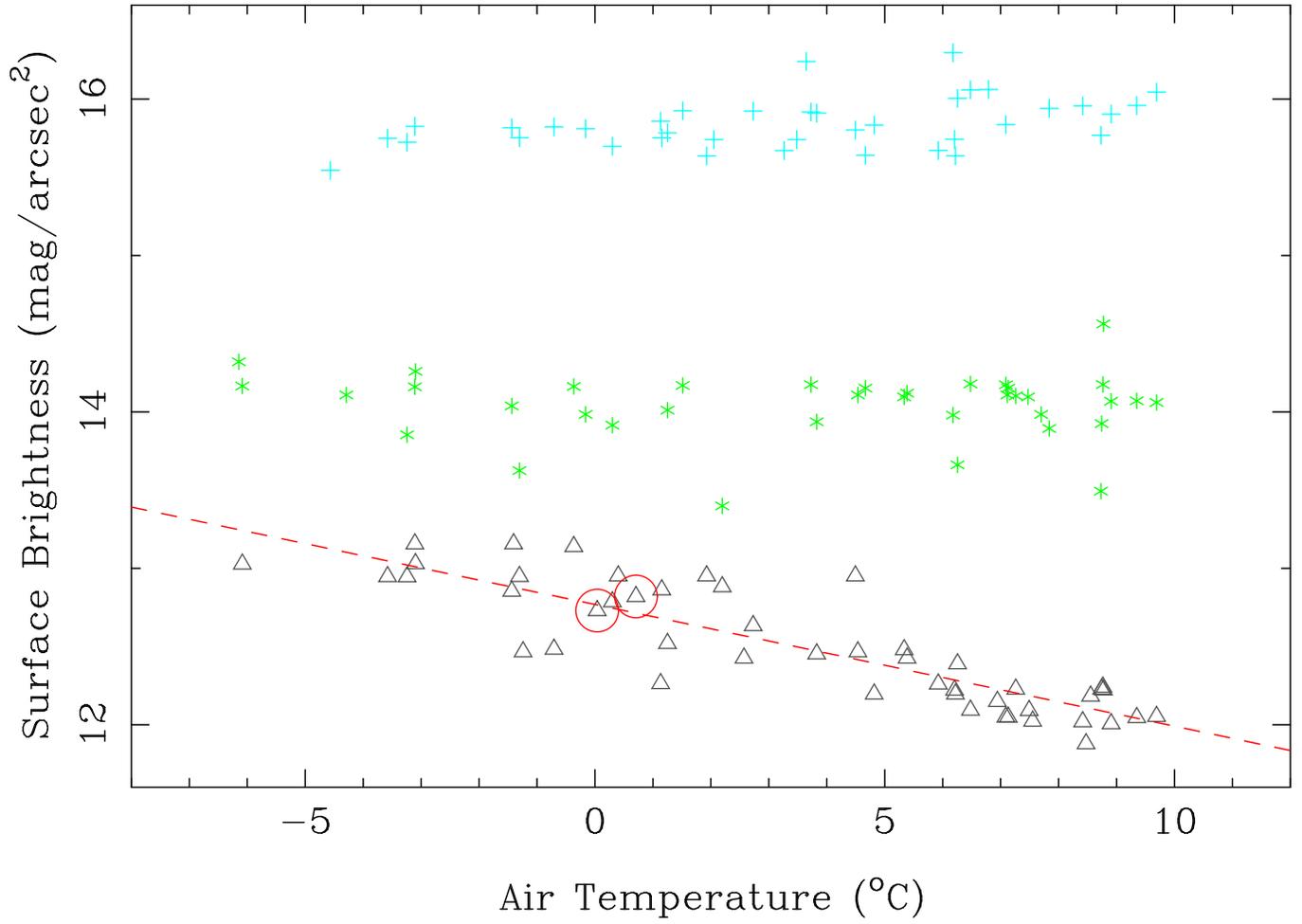}}
\caption{\label{mag_temp} Distribution of the airmass corrected
  surface sky brightness for the $J$, $H$ and $Ks$ bands, averaged for
  each night, along the average night temperature. Only for the
  $Ks$-band a clear correlation between both parameters is seen. The
  dashed line represent the best linear regression fit over the
  $Ks$-band data. The encircle data indicate the data taken with the
  1.23m Telescope.}
  \end{figure}

To investigate the actual dependency of the $Ks$-band sky-brightness with the
temperature we derived the average air temperature measured by the
meteorological station for the nights with measured sky-brightness, comparing
both parameters. In Figure \ref{mag_temp} it is shown the distribution of the
average, air-mass corrected, sky brightness for the $J$, $H$ and $Ks$-band
along the average temperature for the considered nights. As expected there is
a clear dependecy of the sky brightness with the temperature (with a
correlation coefficient of $r=$0.88), only for the $Ks$-band data. A linear
regression derives an increase of $\sim$0.08 mag in the sky brightness per
1$\degree$ increase in temperature. The temperature at Calar Alto presents a
seasonal sinusoudal pattern, with a minimun in winter (December-January) at
$\sim -$8$\degree$ C and a maximum in summer (July-August) at
$\sim$12$\degree$ C. Due to this thermal variance it should be possible to find
differences in the sky brightness of $\sim$2 mag between the coldest nights
in winter and the hottest in summer.

It has been long suspected that the thermal background in the 3.5m telescope
is larger than in rest of the telescopes at Calar Alto, due to its larger
structure. However this claim has never been tested. Although we do not have a
similar sample of sky-backgrounds measured in the rest of the telescopes, the
fact that the $Ks$-band background correlates so tightly with the atmospheric
temperature make us to suspect that this claim is not correct. Indeed, the two
estimations of the sky-brightness taken at the 1.23m, marked with two big
circles in Figure \ref{mag_temp} lie exactly in between the remaining data,
taken at the 3.5m telescope.

\subsubsection{Comparison with other astronomical sites}

As indicated in the introduction the night sky-brightness is a fundamental
parameter to estimate the quality of a certain astronomical site. A fair
comparison of this parameter among different observatories is the best method
to establish their real status. In order to perform this exercise we collected
all the published estimations of the night sky-brightness available in the
literature and in the observatories webpages at different sites. The result
from this collection is listed in Table \ref{tab_other}, including the name of
the observatory where the data were obtained, the kind of measurement listed
(average or darkest sky brightness), the measured sky-brigthness for the $J$,
$H$ and $Ks$-bands and the reference from where the data were extracted.

To perform a fair comparison between the sky brightness at these
different sites and Calar Alto it is required to use the same kind of
measurement. For the darkest sky brightness we will use the
corresponding values listed in Table \ref{tab_zenit}, and for the
average ones we will use those listed in Table \ref{tab_med}. In some
cases, like in Cerron Pachon and Mt.Graham, it is not clearly stated
in the references if they correspond to the average or the darkest sky
brightness. We adopted the less advantage situation for Calar Alto for
the purposed comparison and consider their measurements as the
average (that is always brighter) instead of the darkest.

Regarding the darkest sky-brightness it is clear that for the $J$ (16.96 mag)
and $H$-bands (14.98 mag) Calar Alto can be quoted among the darkest
astronomical sites in the world (or at least the ones listed here). For the
$Ks$ band (13.55 mag) its sky-brightness is clearly comparible with most of
the astronomical sites, apart from Mauna Kea, which is, by far, the darkest
place for this considered wavelength range. This is expected since this site
is located by far at the highest altitude, as it is seeing in Table
\ref{tab_other}, with the most tenuous atmosphere, and therefore the thermal
background there is much lower than in any of the other sites.

For the average sky-brightness we can only compare with the published
sky-brightnesses from el Roque de los Muchachos observatory at La Palma island
and Mauna Kea. In the $J$-band Calar Alto seems to be darker than La Palma or
Mauna Kea by $\sim$0.3 mag and in the $H$-band the three places seem to be
equally dark. On the other hand, in the $Ks$-band Calar Also is slighly
brighter than La Palma, by $\sim$0.2 mag, and both places are much brighter
than Mauna Kea, by $\sim$1 mag, as expected. However, considering the
inhomogeneity of the collected data and the lack of details of how they were
derived we consider that only the differences found with Mauna Kea in the
$Ks$-band are really significative.

\subsection{The seeing measured at the detector}
\label{ana_seeing}

In Paper I we published the median site seeing at the Calar Alto observatory
estimated from a continuous night monitoring of the atmospheric seeing
performed with a DIMM monitor since August 2004 (RoboDIMM, PI:
J.Aceituno). However it is well known that the site seeing is not in most
cases the actual seeing measured in the detector when an astronomical image is
taken. There are several reasons for this effect, being the stronger ones:

\begin{itemize}

\item  The Dome seeing. The turbulence created by the temperature
differences between the air inside the dome and outside the dome creates a
local seeing. For that reason modern telescopes are build with smaller domes
or domes that open wider than tradditional telescopes. In this regards the
large domes of the equatorial mounted telescopes at Calar Alto are clearly a
handicap which in principle could create bad dome seeings. In the last years
there has been a large upgrade in the domes at Calar Alto. New windows were
opened in the dome and large fan systems were installed to create laminar
air-flows that cool down the dome and reduce the turbulence, in order to
limit the effects on the seeing.

\item The mirror seeing. The differences between the temperature on the dome
 and the surface of the primary mirror creates turbulence in its vicinity that
 alters the measured seeing. Due to its larger thermal capacity, mirrors are
 cooled down slower than the dome, and therefore this effect may affect a
 sustantial fraction of the astronomical night. To reduce this effect it was
 installed a cooling system that pumps evaporated nitrogen to the primary
 mirror while the telescope is parked during the day, reducing the thermal
 gradient and its effect over the seeing.

\item Optical problems with the telescope, like bad focusing, misaligments of
  the primary and secondary mirror (or prime-focus instrument), astigmatig
  effects.

\item Mechanical or electronical problems like bad balancing (that creates
  vibrations and fleasures), tracking and/or guiding problems.

\end{itemize}

For all these reasons it is interesting to estimate the seeing measured in the
astronomical images and compare it with the site seeing to estimate the
quality not only of the observatory but also of the telescopes installed on
it. 

For doing so we used the seeing estimated for each frame of the full ALHAMBRA
survey dataset, described in Section \ref{alh}. It is known that the seeing
presents a strong wavelength dependency, being smaller in the redder bands than
in the bluer ones. To perform a fair comparison with the site seeing measured
by the DIMM at the $V$-band it is required to transform the estimated seeing
at each NIR band to the former one. For doing so we adopted the
transformation derived by Sarazin \& Roddier (1990):

$$\rm  Seeing_{\lambda} = Seeing_{\lambda_0} (\frac{\lambda}{\lambda_0})^{-0.2}$$

\begin{figure}
\resizebox{\hsize}{!}
{\includegraphics[width=\hsize,angle=270]{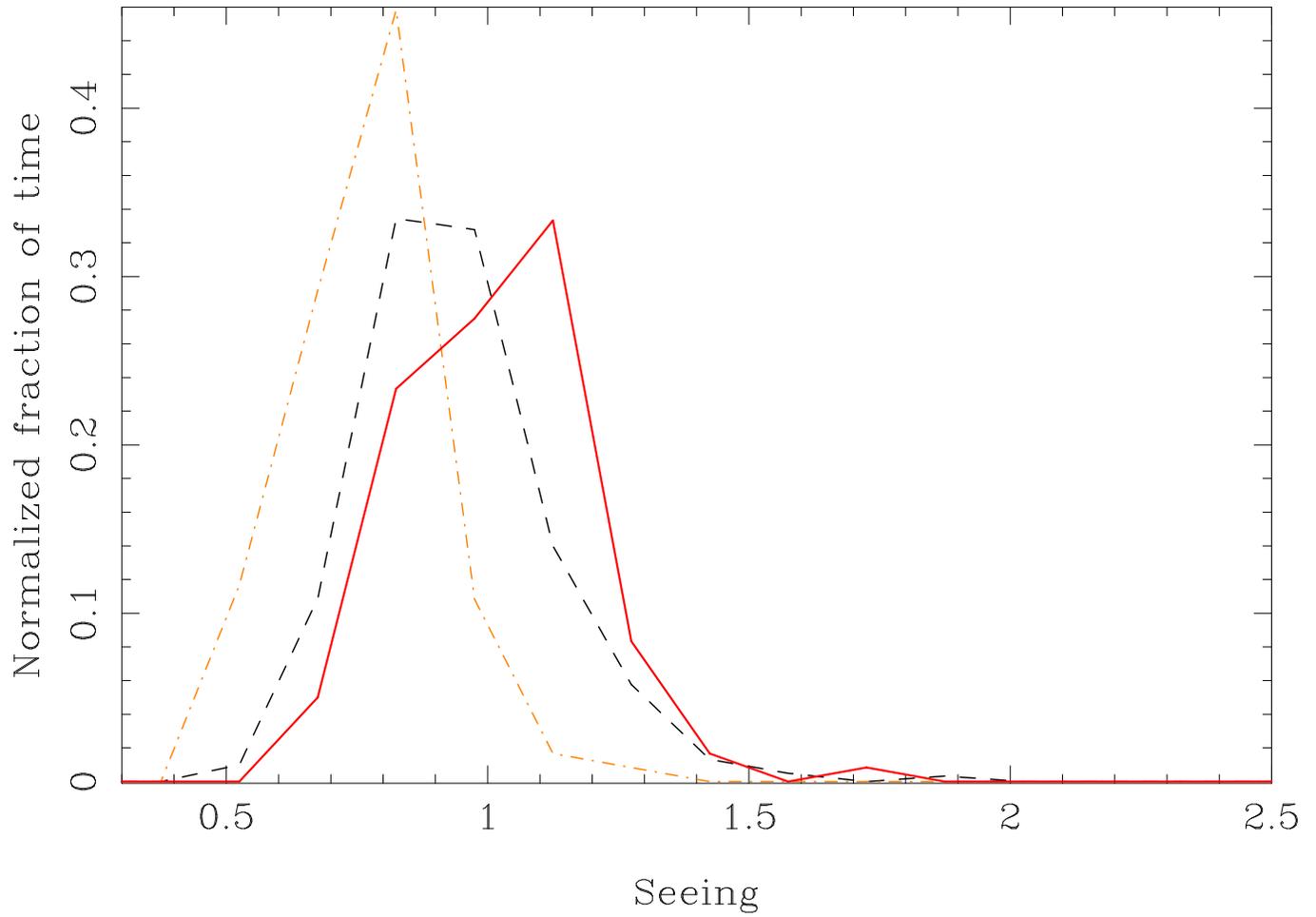}
}
  \caption{\label{seeing} 
Distribution of the telescope seeing at the V-band derived from the analysis of the
~7311 NIR individual images from the ALHAMBRA survey (red solid line)
corresponding to the time period between August 2004 and March 2008, compared
with a similar distribution of the atmospheric seeing, derived from the DIMM
monitor in the same time period (black dashed line). The orange dash-dotted
line shows the distribution of the original seeing measured in the $Ks$-band
images. 
}
  \end{figure}

After applying this transformation its was obtained a final sample of 7311
individual seeing estimations on the detector at the $V$-band that can be
directly compared with the site seeing measured by the seeing monitor. Figure
\ref{seeing} shows the distribution of the seeing measured directly at the
detector together with the distribution of the site seeing measured by the
DIMM monitor in the same time period. As expected the former distribution has
an offset towards larger seeings than the previous one, but there is no
evidence of a broader tail. This indicates that although the telescope seeing
is worst than the site one, there is little evidence of disastrous seeings.
The stronger contribution to the seeing at the detector is site one, and not
the remaining effects descrived before.

The median seeing measured at the detector is $\sim$1.0$\arcsec$ (converted to
the V-band), while the site seeing is $\sim$0.9$\arcsec$ for the same time
period, with a standard deviation of $\sim$0.2$\arcsec$ in both
cases. Therefore, the added degradation of the seeing due to the technical
characteristics of the telescope/instrument is statistically restricted to
$\sim$10\% of the site seeing value. For comparison purposes the measured
seeing distribution in the $Ks$-band is also shown in Fig. \ref{seeing}
(dash-dotted line). The median seeing in this band is $\sim$0.77$\arcsec$,
with a standard deviation of $\sim$0.13$\arcsec$. A $\sim$50\% of the nights
the seeing is lower than this typical value.

\subsection{The Fraction of useful time}
\label{ana_seeing}

A fundamental parameter to estimate the quality of a certain astronomical site
is the fraction of time that it is possible to observe. This concept is always
relative to the instrumentation/telescopes installed in the considered
observatory. For example, radio telescopes can operate both during night and
day time, while optical ones can only operate during the so-called
astronomical night. In this regards the fraction of useful time has to be
computed differently for different telescopes.

The telescopes at Calar Alto were designed to operate in a wide range of
atmospherical conditions. In general they can operate with relative humidities
down to a 95\% and wind gust velocities down to 20 m/s. In certain conditions
of temperatures, cloud coverage and atmospheric preasure they can still operate
with relative humidities as high as a 98\% and wind gust velocities as high as
26 m/s (at least the 3.5m telescope). Therefore the former limit is a
conservative one to define when it is possible to observe. 

The fraction of useful time was estimated for each night using the database of
atmospheric parameters described in section \ref{meteo}, and adopting the
definition of useful time as that time when the telescopes can operate, on the
basis of the conservative limit described before. For the decade comprised in
the dataset it was found that it was possible to observe in a $\sim$72\% of
the night time, and in a $\sim$41\% of the nights the atmospheric conditions
allows to observe for the full night.  This estimation agrees with the
fraction of useful time derived on the basis of the data from the extinction
monitor described in Paper I.

\begin{figure}
\resizebox{\hsize}{!}
{\includegraphics[width=\hsize,angle=270]{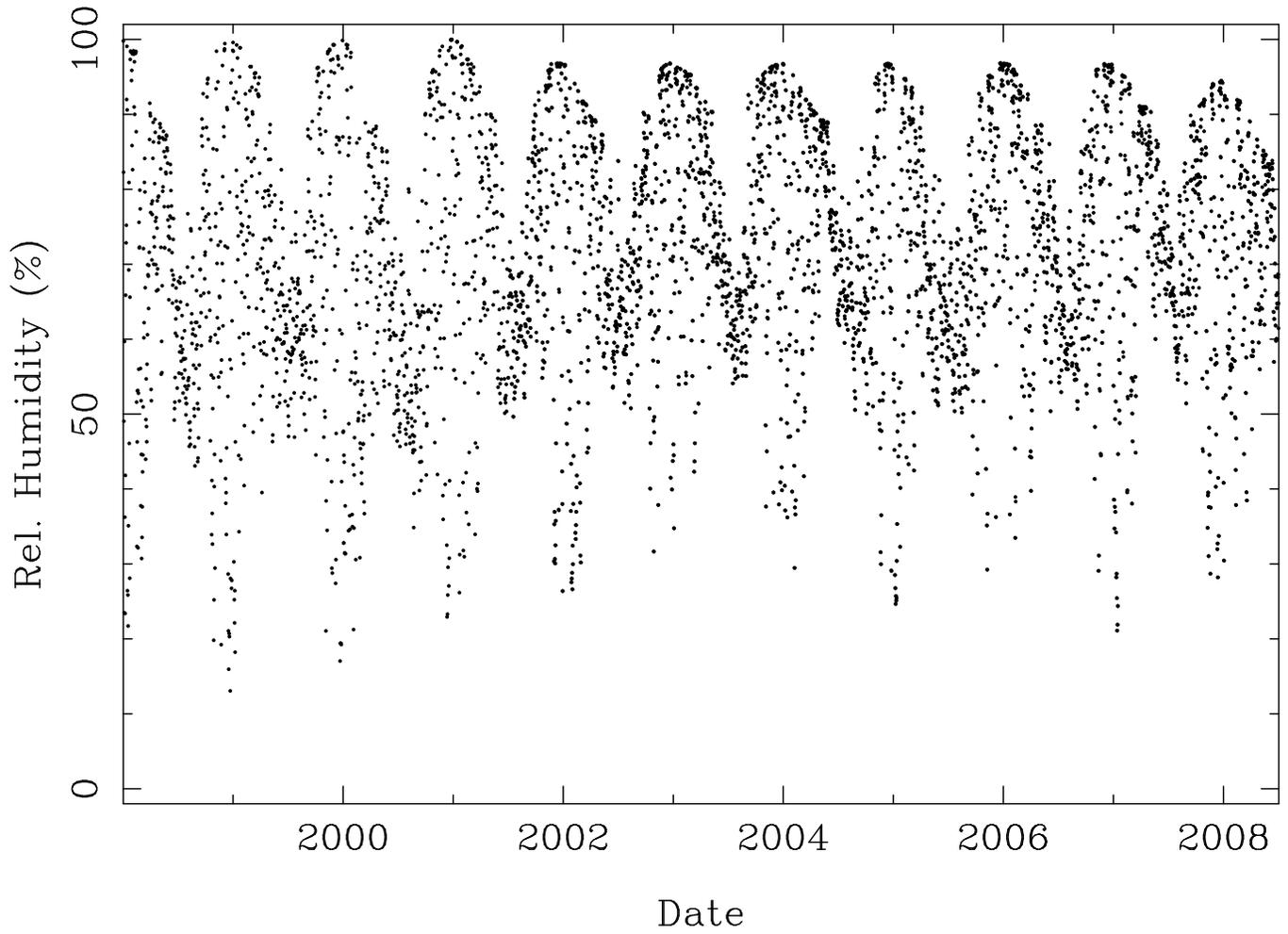}
}
  \caption{\label{humi} 
Distribution of the average relative humidity per
night along the time period between January 1998 and December 2007. 
There is a clear seasonal pattern, with the relative humidity being more
frequently higher and more instable in winter than in summer. The driest
periods are also shown in winter, associated with freezing temperatures. 
}
  \end{figure}

Of the two parameters that define the adopted limit, the humidity is
the more critical one since only in very rare cases a wind gust speed
of 20 m/s is reached. Figure \ref{humi} shows the distribution of
nightly average relative humidity along the time period sampled by the
meteorological station. There is a clear seasonal pattern, with dry
and more stable summers and wet and more instable winters. The peaks
of high relative humidity in winter are due to low altitude clouds and
rain. However, the driest periods are also present in winter, although
less frequently and stable than in summer. These periods are due to
freezing temperatures that drys the atmosphere. This pattern is
reflected in the fraction of useful time, that in summer is $\sim$85\%
in average while in winter drops to a $\sim$55\%.


This estimation has the caveat that in certain conditions it is still possible
to have a covered sky by clouds while the humidity is lower than a 95\%. With
these conditions it will be possible to observe or not depending on the
thickness of the clouds. In these cases the extinction measured by the
extinction monitor is a good estimation of the it is possible to observe or
not. However, as we indicated before there is no big differences between the
fraction of useful times derived by the analysis of the atmospheric conditions
and by the measurements of the extinction.

In order to quantify the presence of clouds it was installed in the
observatory a cloud sensor in October 2006. This monitor determines the
presence of clouds by estimating the difference in temperature between the sky
and the ground, on the basis of the infrared emission of the former one
\footnote{http://www.cyanogen.com/products/cloud\_main.htm}. This instrument is
monitoring this temperature every minute since it become operational, which
provides us with a sample of 466142 individual estimations of this
parameter. Although it does not distinguish between thick and thin clouds, it
is possible to stablish when the sky completely covered, that it is when the
sky is less than 15$\degree$ C colder than the ground. By impossing this
additional limitation to our adopted definition of useful time its actual
fraction drops to a $\sim$68\%. It is interesting to note here that this
fraction is very similar to the fraction of useful time derived directly from
the observations of the ALHAMBRA survey (Section \ref{alh}).

We conclude that the fraction of nights with low humidity and calm wind
and covered by clouds is rather small.

\section{Conclusions}
\label{conc}

In this article we have continued the characterization of the main properties
of the night-sky at the Calar Alto observatory that we started in Paper
I. These properties were compared with similar ones of other different
astronomical sites. The main results of this article can be summarized in the
following points:

\begin{itemize}

\item The night sky brightness at the near-infrared presents a strong
  airmass dependency in the $J$ and $H$ band, and a mild one in the
  $Ks$-band. These dependencies can be modelled and corrected (or
  predicted) for any night and airmass by assuming that they are due
  to airmass dependency of the OH lines emission. Obviously, $J$ and
  $H$-band observations are recomended to be performed near the
  zenith, while at the $Ks$-band this is less critical. This result
  should be taken into account when defining observing strategies in
  the future.

\item There is a seasonal pattern in the $Ks$-band sky brightness, which shows
  an average difference of $\sim$0.8 mag from summer, when is brighter, to
  winter, when is darker. This pattern is due to the strong dependecy of the
  sky-brightness in this band with the atmospheric temperature, which may
  indicate that the contribution from the thermal background emission is the
  dominant component in this band.

\item The brightness of the night sky at Calar Alto is similar or darker than
  any in other major astronomical site in the $J$ and $H$ bands, including
  both Paranal and Mauna Kea. In the $Ks$-band all the astronomical sites are
  very similar apart from Mauna Kea, which is clearly darker (by $\sim$1
  mag). Indeed, Calar Alto has a similar sky brightness than Paranal or La
  Palma at this wavelength range.

\item The seeing on the detector is only slightly larger than the site seeing
  measured by the DIMM, outside the dome (at least for the 3.5m
  telescope). The effects different than the purely atmospheric ones
  produce a statistical increase of a $\sim$10\% in the measured seeing on the
  detector compared with the site one.

\item The fraction of useful time at the observatory is of the order of
  $\sim$70\%, with $\sim$40\% of the nights 100\% completely useful. These
  fractions are better during Summer and worst during Winter.

\end{itemize}

We finally conclude that our additional analysis of the astronomical conditions
at Calar Alto agree with our previous results presented in Paper I. The new
data strength our previous conclusion that it is a good astronomical site,
similar in many aspects to places where there are 10m-like telescopes under
operation or construction.


For both reasons we consider that this observatory is a good candidate
for the location of future large aperture optical/NIR telescopes.

\section{Acknowledgments}

  SFS thanks the Spanish Plan Nacional de Astronom\'\i a program
  AYA2005-09413-C02-02, of the Spanish Ministery of Education and Science and
  the Plan Andaluz de Investigaci\'on of Junta de Andaluc\'{\i}a as research
  group FQM322.
  
  We acknowledge Dr. M.Moles, PI of the ALHAMBRA survey, for allowing us the
  access to the data of this project.

\newpage

\begin{table}
\begin{center}
\caption{Log of the data-set per night}
\label{tab_data}
\begin{tabular}{llrrl}\hline
\tableline\tableline
Date(s)     & Telescope & Instrument & Bands  & Number of frames\\
\tableline
30/01/08    & 1.23m     & Magic      & J,H,Ks,K,Km & 24 per band\\
31/01/08    & 1.23m     & Magic      & J,H,Ks,K,Km & 24 per band\\
ALHAMBRA(1) & 3.5m      & Omega2000  & J,H,Ks & 17\\
ALHAMBRA(2) & 3.5m      & Omega2000  & J,H,Ks & 1954,2394,2963 \\
\tableline
\end{tabular}

(1) 7 nights of observations in August 2004 from the ALHAMBRA survey.
    Analysis based on the reduced and combined frames.\\
(2) 75 nights of observations between August 2004 and March 2008 from the
    ALHAMBRA survey. Analysis based on the individual frames.\\

\end{center}
\end{table}

\begin{table}
\begin{center}
\caption{Typical values of the night-sky surface brightness at Calar Alto in the NIR}
\label{tab_med}
\begin{tabular}{llllll}\hline
\tableline\tableline
Description &        J   &    H       &    Ks     &    K    &   Km  \\  
\tableline
\multicolumn{6}{c}{Calar Alto, 1.23m Telescope, MAGIC, 30/01/08}\\
\hline
median	    &   16.45	 &   14.36    &	  12.57   & 11.97   &   12.78 \\	
brightest   &   16.18    &   14.18    &   11.49   & 11.88   &   12.69 \\
darkest     &   16.69    &   14.54    &   12.73   & 12.15   &   12.88 \\
\hline
\multicolumn{6}{c}{Calar Alto, 1.23m Telescope, MAGIC, 31/01/08}\\
\hline
median	    & 	15.53	 &   14.03    &	  12.54   &  12.02  &   12.76 \\
brightest   &   14.91	 &   13.39    &   12.38   &  11.81  &   12.54 \\
darkest	    &   16.52	 &   14.98    &   12.82   &  12.34  &   13.04 \\
\hline
\multicolumn{6}{c}{Calar Alto, 3.5m Telescope, Omega2000, ALHAMBRA, August 2004}\\
\hline
median      &	16.04	 &   14.04    &   12.20   & $-$ & $-$ \\
brightest   &   15.45	 &   13.45    &   11.93   & $-$ & $-$ \\
darkest	    &   16.57	 &   14.80    &   12.70   & $-$ & $-$ \\
\hline
\multicolumn{6}{c}{Calar Alto, 3.5m Telescope, Omega2000, ALHAMBRA, Full sample}\\
\hline
median      &    15.64   &   13.76    &   12.28   & $-$ & $-$ \\
darkest     &    16.96   &   14.98    &   13.55   & $-$ & $-$ \\      
\tableline
\end{tabular}
\end{center}
\end{table}


\begin{table}
\begin{center}
\caption{The night-sky surface brightness at Calar Alto in the NIR at the Zenith}
\label{tab_zenit}
\begin{tabular}{llll}\hline
\tableline\tableline
Description     &        J   &    H       &    Ks      \\  
\tableline
\hline
Airmass$<$1.05  &     15.92  &	 13.88    &   12.49     \\
\hline
Airmass corrected&    15.95  &   13.99    &   12.39     \\       
Winter          &     15.92  &   13.99    &   12.89     \\       
Summer          &     15.97  &   14.02    &   12.07     \\
\tableline
\end{tabular}
\end{center}
\end{table}

\begin{table}
\begin{center}
\caption{The night-sky surface brightness in other observatories}
\label{tab_other}
\begin{tabular}{lrcllll}\hline
\tableline\tableline
Description    & Height (m) & Type &        J   &    H       &    Ks     &    Reference\\  
\tableline
La Palma       & 2500     & Average &      15.5  &   14.0     &    12.6   &   1 \& 2 \\
Paranal        & 2635     & Darkest &      16.5  &   14.4     &    13.0   &   3 \& J.Cuby et al. (2000)\\
	       & ``       & Darkest &      $-$   &   $-$      &    13.2   &    Kervella et al. (2006)\\
               & ``       & Darkest &      $-$   &   $-$      &    13.5   &    Boehnhardt et al. (2003) \\
Cerro Pachon   & 2200     &   ?     &      16.0  &   13.9     &    13.5   &  4 \\
Mt.Graham      & 1926     &   ?     &      $-$   &   $-$      &    13.5   &    5 \\
Mauna Kea      & 4200     & Darkest &      16.75 &   14.75    &    14.75  &   6 \\
               & ``       & Average &      15.6  &   14.0     &    13.4   &   7 \\		       
Mt.Hamilton    & 1283     &   ?     &      16.0  &   14.0     &    13.0   &   8 \\
Kitt Peak      & 2096     &   ?     &      15.7  &   13.9     &    13.1   &   9 \\
Anglo Australian Obs. & 1164  &  ?  &      15.7  &   14.1     &    13.5   &   10 \\  
\hline
\multicolumn{7}{l}{(1) F. Ghinassi: http://www.tng.iac.es/instruments/nics/imaging.html\#zpoints}\\
\multicolumn{7}{l}{(2) M.Pedani: http://www.tng.iac.es/info/la\_palma\_sky.html}\\
\multicolumn{7}{l}{(3) http://www.eso.org/gen\-fac/pubs/astclim/paranal/skybackground/}\\
\multicolumn{7}{l}{(4) http://gemini.fcaglp.unlp.edu.ar/sciops/ObsProcess/obsConstraints/obsConstraints.html}\\
\multicolumn{7}{l}{(5) http://www.bo.astro.it/$\sim$ciliegi/astro/nirvana/DOCUMENTS/doc\_nirvana\_V1.2.pdf}\\
\multicolumn{7}{l}{(6) http://www.jach.hawaii.edu/UKIRT/astronomy/sky/skies.html}\\
\multicolumn{7}{l}{(7) http://casu.ast.cam.ac.uk/surveys\-projects/wfcam}\\
\multicolumn{7}{l}{(8) http://mthamilton.ucolick.org/techdocs/instruments/ircal/ircal\_detector.html }\\
\multicolumn{7}{l}{(9) http://www.noao.edu/kpno/manuals/flmn/flmn.html}\\
\multicolumn{7}{l}{(10) http://www.aao.gov.au/AAO/iris2/iris2\_overview.html\#Imaging sensitivities}\\
\end{tabular}
\end{center}
\end{table}


\newpage

 \label{lastpage}


\end{document}